# Bound states in the continuum and long-range coupling of polaritons in hexagonal boron nitride nanoresonators

Harsh Gupta[1, 2, *], Giacomo Venturi[3, 4], Tatiana Contino[1,2], Eli Janzen[5], James H. Edgar[5], Francesco de Angelis[1], Andrea Toma[1], Antonio Ambrosio[3], Michele Tamagnone[1, *]

1) Istituto Italiano di Tecnologia, via Morego 30, 16163 Genova, Italy

2) Dipartimento di Chimica e Chimica Industriale, Università degli Studi di Genova, via Balbi 5, 16126 Genova, Italy

3) Center for Nano Science and Technology, Fondazione Istituto Italiano di Tecnologia, Milan, Italy

4) Dipartimento di Fisica, Politecnico Milano, Piazza Leonardo Da Vinci 32, Milano 20133, Italy

5) Tim Taylor Department of Chemical Engineering, Kansas State University, Manhattan, KS, 66506, USA

* Corresponding Authors harsh.gupta@iit.it, michele.tamagnone@iit.it

## *Abstract*

**Bound states in the continuum (BICs)** garnered significant for their potential to create new types of nanophotonic devices. Most prior demonstrations were based on arrays of dielectric resonators, which cannot be miniaturized beyond the diffraction limit, reducing the applicability of BICs for advanced functions. Here, we demonstrate BICs and quasi-BICs based on high-quality factor phonon-polariton resonances in isotopically pure h$^{11}$BN and how these states can be supported by periodic arrays of nanoresonators with sizes much smaller than the wavelength. We theoretically illustrate how BICs emerge from the band structure of the arrays and verify both numerically and experimentally the presence of these states and enhanced quality factor. Furthermore, we identify and characterize simultaneously quasi-BICs and bright states. Our method can be generalized to create a large number of optical states and to tune their coupling with the environment, paving the way to miniaturized nanophotonic devices with more advanced functions.

## *Main*

Polaritons, a fascinating phenomenon in the realm of nanophotonics, offer a remarkable capability: subwavelength confinement of light at the nanoscale[1–20]. This phenomenon arises from the strong coupling of photons with collective excitation in matter, such as electronic motion (plasmons) or vibrational modes (phonons)[12,16]. This marriage results in hybrid light-matter modes possessing unique properties, such as subwavelength confinement due to the tight spatial localization of polariton modes, opening doors to a plethora of applications that benefit from their ability to beat the diffraction limit and to enhance light-matter interactions.

Examples include super-resolution imaging techniques[17], chemical and biological sensors and detectors with enhanced sensitivity[1,2], miniaturized nanophotonic devices, improved energy harvesting and photovoltaics, and metamaterials with novel optical properties.

In this study, we investigate the existence of bound states in the continuum in arrays of high-quality phonon polariton resonators using isotopically pure hexagonal boron nitride (h$^{11}$BN). The term "Bound State in the Continuum" (BIC) was first coined by von Neumann and Wigner[21,22], in 1929. They found that with a carefully engineered potential, the wave function of a particle could remain bound despite the presence of a continuum of (propagating) states having an energy range which includes one of the bound states. In general, there can be resonant states in the continuum, which, however, are coupled to it and radiate (or "leak"), leading to a finite lifetime. Instead, BIC states do not couple or radiate, even if their energy level would allow it.

Because of that, BICs have, in principle, an infinite lifetime and fine tunability of frequency response based on the geometry of resonators, making a fundamental aspect that distinguishes them from the other resonant states that have a finite lifetime. However, the concept of BICs remained a theoretical curiosity because of its non-radiative nature until the concept of experimentally produced longer lifetime (not infinite) quasi-BIC states (q-BICs) with a little radiative nature was introduced in the 1970s[23] when it was realized that they could have practical applications in areas such as photonics and microelectronics[24]. In fact, BICs and q-BICs can occur for light as well, for instance, in nanophotonic resonators, where these states are essentially dark states and quasi-dark states, respectively [24–34]. Most of the existing BIC systems are based on dielectric resonators[25–29], which however, due to the diffraction limit, cannot be miniaturized beyond a fraction of wavelength. We demonstrate here that BICs can be easily extended to hBN polaritons using structures that can be arbitrarily small, which opens many possibilities in terms of design complexity and allows for miniaturized device with the ability to integrate multiple functionalities into a small footprint. We present here the conditions and mechanisms that give rise to these bound states in our system.

## *Theory*

BICs are supported by several types of nanophotonic dielectric structures. Dielectrics with high refractive index (such as silicon in the near infrared) are generally used because they enable efficient light confinement and compact device integration, which makes them ideal for a wide range of applications[25,30]. For this work, we directly extended one of the most common structures: an array of pairs of elliptical resonators where ellipses are rotated by a small angle $\theta$ in opposite directions (see Figure 1a, b). First, the elliptical polaritonic resonators are designed, and then the unit cell formed by the pair of ellipses is optimized. Elliptical-shaped resonators are used here because they are easier to fabricate experimentally and have been used successfully for BICs in other platforms. The first step was covered by previous works[1,10,14,19] which considered disks, ribbons, and ellipses made of hBN. Here, the use of isotopically pure h[11]BN enhances the polariton propagation lengths about three times better than their natural abundance, which significantly increases the quality factor of resonators without requiring any design alterations[8,19]. Since hBN is a wideband insulator (with an indirect band gap of nearly (6.3eV), [35] free carrier losses are negligible, which preserves the quality of its natural hyperbolic phonon polaritons, making it a suitable candidate for this work. The uniaxial response is due to optical phonons with different dispersion for ion core oscillations parallel or perpendicular to the van der Waals layers. In the limit of small momentum, these optical phonons couple strongly with the photon field, creating a strong polaritonic response in two bands, the reststrahlen band 1 (RB1, Type-I, 764 cm$^{-1}$ – 820 cm$^{-1}$) and 2 (RB2, Type II, 1370 cm$^{-1}$ – 1613 cm$^{-1}$), created by out-of-plane and in-plane oscillations respectively[14,19].

For the structures we considered, the momentum of the polaritons is much smaller than the reciprocal lattice of hBN, Thus, the non-locality of phonon-polaritons can be neglected, and their optical response in the linear limit can be represented by a single oscillator per band in a simple Lorentz model. The permittivity tensor is therefore diagonal with $\varepsilon_{xx} = \varepsilon_{yy} \neq \varepsilon_{zz}$:

$$\varepsilon_{ii} = \varepsilon_{\infty,i}\left(1 - \frac{\omega^2_{\text{LO},i} - \omega^2_{\text{TO},i}}{\omega^2 - i\omega\Gamma_i - \omega^2_{\text{TO},i}}\right) \tag{1}$$

where $i \in \{x, y, z\}$, $\omega = 2\pi f = 2\pi c/\lambda$ is the angular frequency, $c$ is the speed of light, $\lambda$ is the wavelength of the incident radiation, $\varepsilon_{\infty,i}$ is the high-frequency limit of the permittivity, $\varepsilon_i(\text{x})$, and $\varepsilon_\infty(\text{z})$ is the high-frequency limit of the permittivity along x-and z-direction respectively, $\Gamma$ is the resonance linewidth, $\omega_{\text{LO},i}$ and $\omega_{\text{TO},i}$ are the upper and the lower frequency limits of the reststrahlen bands along x- and z-direction, respectively[8].

Inside the reststrahlen bands some of these entries are negative, which is the key for some of the properties of hBN polaritons such as hyperbolicity. Here, however, hyperbolicity is not required, and we simply rely on the existence of surface waves that, when confined to the chosen elliptical shape, create the resonant modes. Because of this, the following steps can be extended to polaritons in other 2D and van der Waals materials. Note that the presence of the negative permittivity that allows these polaritonic structures to be miniaturized essentially without limitation, and they will always resonate in the restrhalen bands. The elliptical resonator considered here, has two main resonances, both are dipolar and lie in the RB2 band: one along the major axis and the other (at higher frequency) along the minor axis.

We consider an array of aligned h[11]BN ellipses on a CaF$_2$ substrate, so that each unit cell contains a single elliptical resonator (Figure 1a). Due to the interelement coupling, each original mode of the resonators generates a mode in the band structure of the array[36] (Figure 1e). This is due to the fact that the resonant frequency depends on the phase difference $\phi$ between each resonator and its nearest neighbour. If the periodicity of the array is $P$ then the phase difference can be associated with an in-plane wavevector $k = \phi/P$. The band diagram them simply relates $\omega$ to $k$ for each mode.

A necessary (but insufficient) condition for a mode to be bright is to be inside of the light cone delimited by $\omega > c|k|/n$ where $c$ is the speed of light and $n$ is the biggest among the refractive indexes of substrate and superstrate. Conversely, modes outside the light cone are always dark. In our case, because we work with polaritonic resonators that are deeply subwavelength, we will design structures with very small periodicity so it easy to ensure that there are points in the band structure outside the light cone, and therefore dark modes. Two notable points of the band structure are (Figure 1d, 1e):

- The $\Gamma$ point for $\phi = 0$, meaning $k = 0$, the centre of the Brillouin zone which can couple with normally incident light. We will refer to this mode as "bright mode".
- The $X$ point for $\phi = \pi$, meaning $k = \pi/P$, the darkest mode at the edge of the Brillouin zone. We will refer to this mode as "dark mode" or BIC.

We then progressively break the symmetry between adjacent ellipses by rotating them by a small angle $\theta$ in alternating directions (Figure 1b, 1c). Because of this, the new unit cell has periodicity twice of the initial cell (x-periodicity $P_x = 2P$; y-periodicity $P_y$). For this cell, the Brillouin zone is halved, and part of the band structure is folded, forming a new branch. The old dark mode $X$ is renamed $X'$ (Figure 1e) and is folded at the $\Gamma$ point and can now couple to normally incident light (but notice that the polarization must be orthogonal compared to the corresponding bright mode). The coupling strength increases monotonically with the magnitude of the perturbation, so that the new mode is quasi-dark, i.e., quasi-BIC.

Coupling changes the band structure from a flat line and ultimately for the frequency shift between BIC and bright modes. This phenomenon occurs in general for any mode of the resonators. Since we focus here on the two aforementioned dipolar modes of the ellipses, we identify and study these 4 modes in the ellipse pair:

- **Ba**: Bright mode associated with major axis (a) mode.
- **Da**: (quasi-)Dark mode (or q-BIC) associated with major axis mode.
- **Bb**: Bright mode associated with minor axis (b) mode.
- **Db**: (quasi-)Dark mode (or q-BIC) associated with minor axis mode.

One of the key motivations to study these modes is the possibility of increasing the resonance quality factor due to the suppression of radiation losses, which are expected to be approximately proportional to $\theta^2$ and are studied in more detail in the next sections.

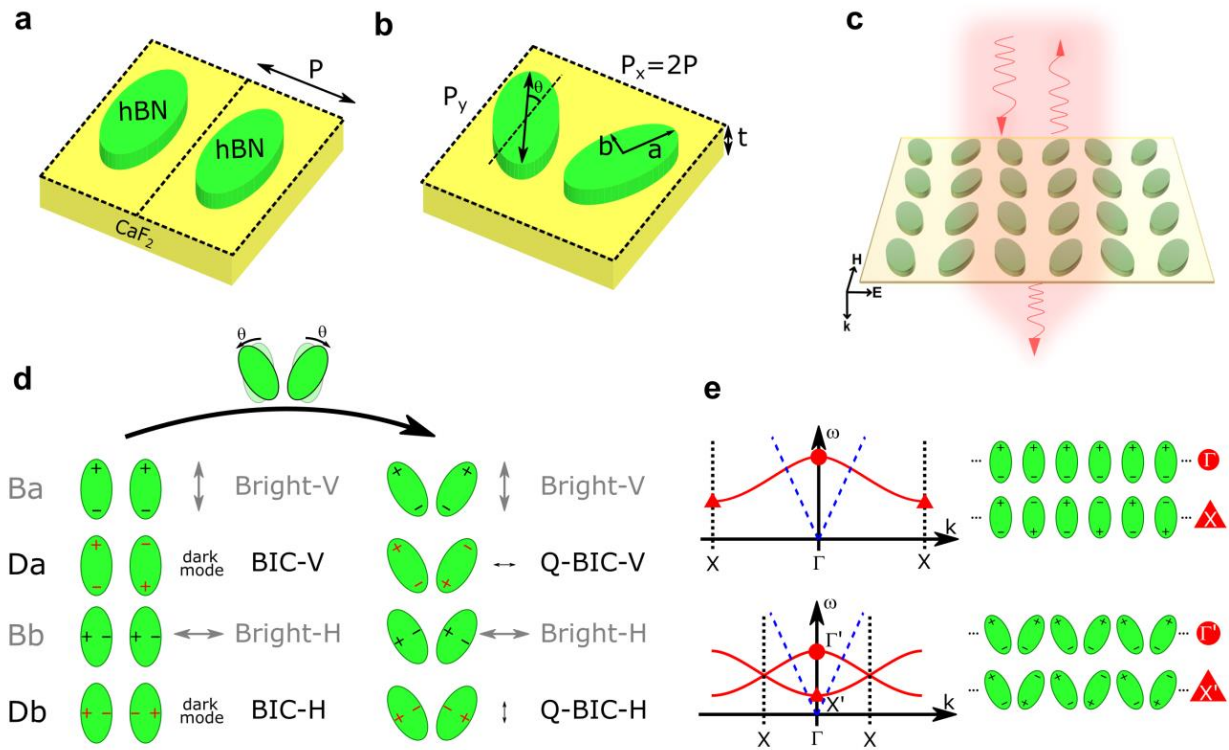

**Figure 1. Unit cell and band structure of the BIC resonators**. **(a)** A pair of symmetrical elliptical hBN resonators on the CaF2 substrate. Dash lines delimit the single-resonator unit cell. **(b)** Unit cell of the BIC system formed by two elliptical resonators rotated by $\theta = 25°$ in opposite directions. For both panels, major axis $2a$ and minor axis $2b$ are 1.025 μm and 514 nm respectively (quantities taken directly from the fabricated devices shown below), periodicity is P = 0.9 μm, $P_x = P_y$ =1.8 μm. hBN thickness is 50 nm. **(c)** Metasurface formed by an array of cells with incident and scattered light. **(d)** Bright and dark (BIC) modes supported by the unit cells have different resonances due to the coupling between resonators. When the rotation $\theta$ is introduced (in the limit of small $\theta$) the previous BIC modes become q-BIC since they can now weakly couple to incident far-field light. However, the polarization coupled with far-field light must have orthogonal polarization with respect to the corresponding bright mode. **(e)** Schematic sketch of structures for symmetrical and symmetry-broken cells (the light cone is represented in dashed blue lines). For the symmetric system, the central point (Γ) represents the bright mode for normal incidence, while the band edge case (X) represents the darkest mode, with dipoles having alternating signs. Breaking the symmetry halves the width of the band structure and folds it so that the previous X mode (X') is now folded into the light cone and weakly couples to normal incidence light of orthogonal polarization. Coupling is responsible for the band structure to be different from a flat line and ultimately for the frequency shift between BIC and bright modes.

## *Numerical simulations*

We numerically simulated the resonator to select the physical dimensions to be compatible with the requirements above, with the capabilities of our nanofabrication systems and to show a significant resonant frequency separation to facilitate the experiment. Selected parameters are reported in Figure 2. We simulated transmission for both incident polarizations along H and V (Figure 2a-c) for values of the angle $\theta$ from 0 to 30°. Each of the four considered modes creates an absorption dip in the transmission spectrum. However, while bright modes are not significantly affected by the asymmetry parameter $\theta$, the absorption dips of the q-BIC modes become deeper with larger $\theta$, as expected due to the increased coupling with the far field. Based on these plots, a nominal rotation of $\theta = 25°$ was chosen for the final experiment and will be used in the remainder of this work. Field plots (Figure 2d-g) are obtained exciting the system at normal incidence with the polarization and frequency of each of the mode, and they match the theoretical model discussed earlier. For the long-axis modes, we find that the spatial distribution of the quasi-BIC mode is confined to a smaller region than the BICs in the symmetric system. This explains the fact that the resonance frequency of that q-BIC mode is redshifted for larger angles (Figure 2b)

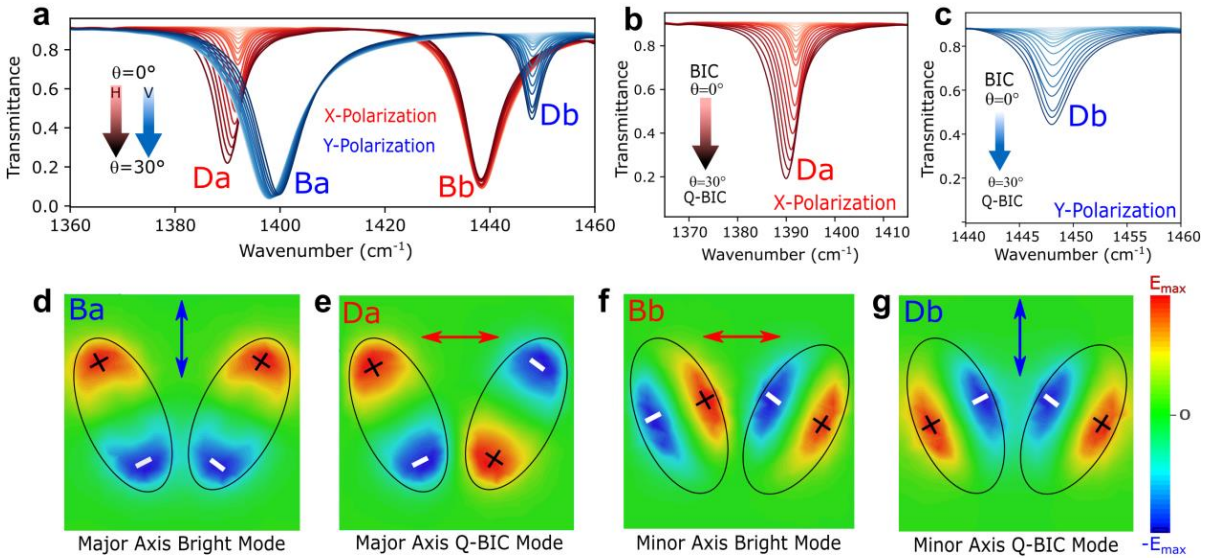

**Figure 2. Simulations and electric field plots. (a)** Numerically simulated transmission spectra for different values of the asymmetry parameter θ in the range 0° to 30° and for both incident polarization x and y. **(b, c)** Magnification of **(a)** around the two q-BIC modes Da and Db. Unlike the bright modes (which are marginally affected by θ), the absorption dip of the q-BIC modes increases for increasing θ, and vanishes for θ=0. **(d-g)** Numerically simulated electric field plots (component $E_z$ immediately above the resonators) with same dimensions as Figure 1, for all the considered modes using an incident plane wave with polarization and frequency matching each of the targeted modes. The sign of the plot correlates with the sign of the induced charge density due to the ion core displacement and provides an excellent match to the theoretical model in Figure 1.

Figure 3a. represents the phenomenological quality factor of the simulated resonance, defined as the ratio of the resonant frequency to the FWHM (full width at half maximum). As expected, the Q factor is maximized for low values of the perturbation parameter $\theta$. The actual quality factor can be slightly different due to the saturation effect of the spectrum when the dip is close to the full extinction

condition. A higher value of Q leads to a higher Purcell enhancement which is useful for applications involving light-matter interactions, such as sensing and photodetectors.

The hBN thickness is also an important design parameter. Figures 3b-d illustrate the behavior of the system for different hBN thicknesses. Larger thicknesses lead to a blueshift of all resonances (Figure 3c-d) and, more importantly, to a larger separation between q-BIC and bright modes (Figure 3b). This is due to an increased coupling between resonators. Therefore, we selected an initial flake with a thickness of approximately 50 nm and adjusted the design of the resonators accordingly.

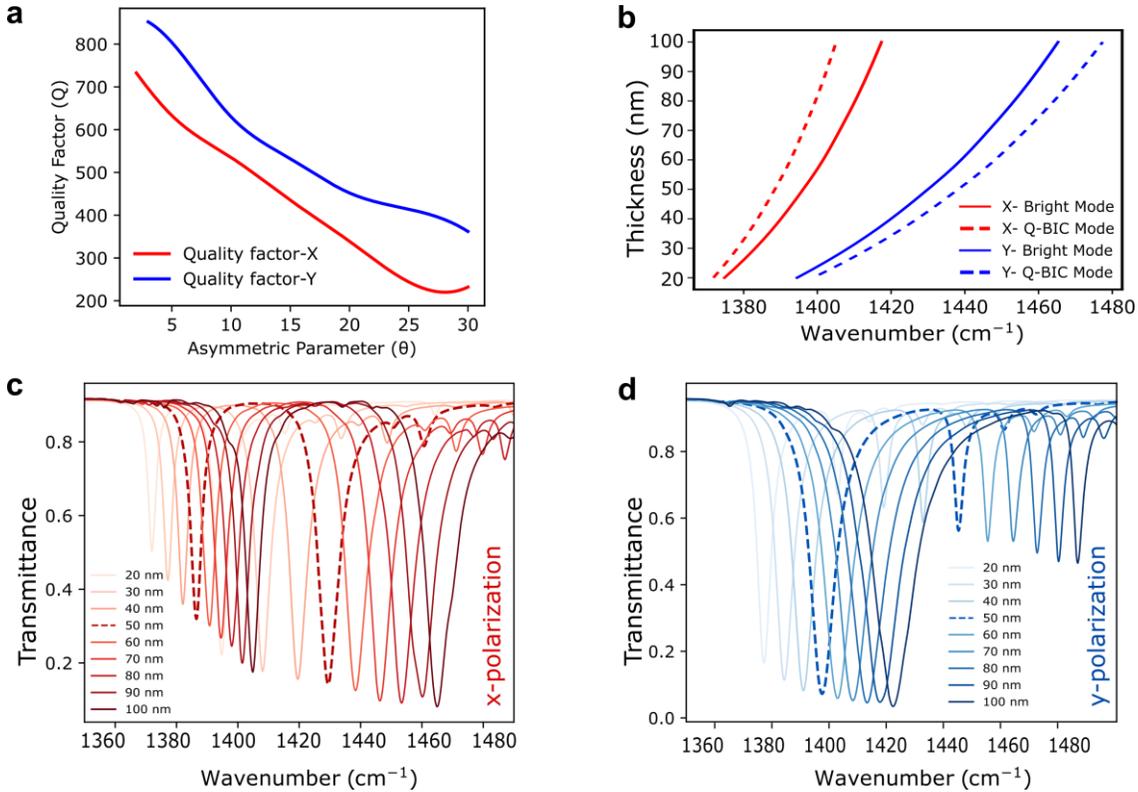

**Figure 3. Q factor and thickness dependence of q-BIC states in hBN. (a)** Dependence of the Q-factor of the q-BIC state on the asymmetry parameter (θ) along x and y-polarization (red) and (blue), respectively. **(b)** Parametric simulations with thickness variation from 20 nm to 100 nm (steps of 10 nm) were performed keeping $\theta = 25°$. Variation in resonant frequencies as a function of the thickness of hBN resonators. Larger thickness leads to a blueshift in the resonances and to a larger separation between q-BIC and bright modes. This is due to increased coupling between resonators. **(c,d)** Simulated transmission spectra for both polarizations, as a function of the thickness. Dashed lines mark the spectra in both plots for the target hBN thickness (50 nm).

## *Experiments*

The h[11]BN was first exfoliated by scotch tape over a $CaF_2$ substrate and then fabricated with a simple e-beam lithography step followed by reactive ion etching (RIE) as explained in the Methods (Figure 4a-c).
To verify the presence of the q-BIC and bright modes, we measured the far field transmittance using a FITR (Fourier-transform infrared spectroscopy) system coupled with a microscope (Methods, Figure 4e).
The experimental FTIR curves (Figure 4e) match well with the simulations (Figure 4d), except for a broadening and shortening of the dips in the experiment due to imperfections and disorder in the fabricated resonators. The fact that the incident light is focused instead of a plane wave also contributes to these non-idealities. Notice that the resonances are very clean and only appear for the correct polarization as predicted from the theory. This experiment already confirms that we successfully observed q-BIC states in a polaritonic system. Table 1 reports the measured resonances and compares them to the simulations:

| Mode   | Freq (SIM), cm$^{-1}$ | Freq (FTIR), cm$^{-1}$ | $Q_P$ (SIM) | $Q_P$ (FTIR) |
|--------|-----------------------|------------------------|-------------|--------------|
| **Da** (y) | 1386.9            | 1383                   | 300         | 198          |
| **Ba** (x) | 1393.8            | 1396                   | 106         | 143          |
| **Bb** (x) | 1429.7            | 1427                   | 145         | 142          |
| **Db** (y) | 1438.2            | 1437                   | 366         | 160          |

**Table 1. Simulated and experimental resonances**. Modes are sorted by frequency. Resonance frequency and phenomenological quality factor $Q_P$ are reported.

The table also compares the Q factor of the resonances for experiment and theory. In this case, we use the phenomenological quality factor $Q_P$ defined as[28]:

$$Q_P = \frac{\omega_0}{\Delta\omega} \quad (2)$$

where $\omega_0$ is the resonant frequency of the mode and $\Delta\omega$ is the full width at half-maximum (FWHM) of the same mode evaluated from the transmission spectrum. Note that this quantity may be slightly affected by saturation phenomena for absorption lines that approach zero at their minimum. It is also important to note that the losses in hBN will limit the Q factor, but values close to the theoretical limit can be reached in this way, estimated here to be approximately 800.

For the long-axis modes (**Da, Ba**) the quality factor is enhanced 183% numerically and 38% experimentally with respect to bright modes, while for the short-axis ones (**Db, Bb**) the quality factor is enhanced 152% numerically and 13% experimentally with respect to bright modes. These figures confirm the enhancement discussed above.

The quality factor enhancement is also related to an improved lifetime τ of the polaritonic mode as the relationship between lifetime and Q factor can be expressed mathematically as:

$$\tau = \frac{Q}{\omega_0} \quad (3)$$

The larger value of the quality factor, and therefore a larger lifetime of the polariton, allows for a prolonged light-matter interaction and higher field enhancement, which can increase the performance of devices such as photodetector and biosensing, and can potentially find application also for non-linear optics.

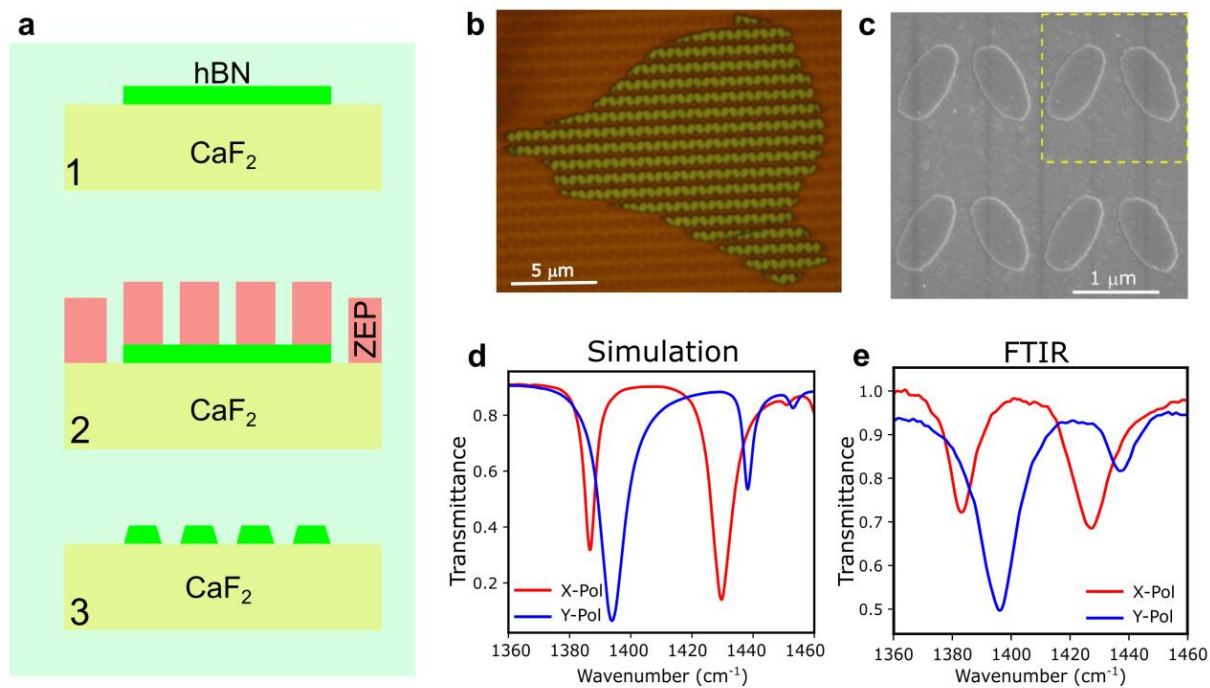

**Figure 4. Schematic description of the nanofabrication and transmittance curve comparison. (a)** Illustration of the nanofabrication of hBN elliptical resonators. **(b)** Optical microscope image of the hBN resonators after lithography. **(c)** Scanning electron microscopy (SEM) imaging of the elliptical hBN resonators with $\theta = 25°$ in which a yellow-colored dashed square represents a unit cell of the array. **(d-e)** Comparison of numerical and experimental transmission spectra. **(d)** Shows the same curves as in Figure 2a for the case $\theta = 25°$. The experimental curves in **(e)** where obtained using a polarization-resolved FTIR system coupled to a microscope (See Methods). The unpolarized trace has been added with a scale factor of 2 to better compare with the polarized ones.

The far field measurements already prove clearly the experimental observation of the BIC states, but do not offer a direct validation concerning the geometry of the modes. To obtain this validation, we used a nanoscale version of FTIR (nano-FTIR) based on a scanning near-field optical microscope (SNOM) of a commercial Neaspec system (Figure 5a). During these measurements, a broadband mid-IR laser was focused under a sharp AFM tip. The tip was scanned with respect to the sample surface and the scattered light was collected by a detector, whose signal (after demodulation with tip oscillations see methods) represents the local near-field of the resonator with a spatial resolution of the order of the tip radius (about 50 nm). An interferometric setup allows the retrieval of the complex spectrum, as in Figure 5b-c.

Figure 5b-c represents the near field response as function of the position on the axes of the ellipses and the wavenumber. For these experiments we first aligned the sample in such a way that the polarization of the incident light would couple with one target q-BIC mode ("**Da**" and "**Db**" in Figure 4b and 4c respectively). We then collected spectra while scanning the tip along the corresponding axis of one of the ellipses (major axis for Figure 5b, minor axis for Figure 5c). Thanks to this arrangement we see the q-BIC mode; but at the same time, we also see a portion of the bright mode. This due to the fact that the bright mode has a much stronger coupling, and can be excited due to the experimental imperfections of the system. Incidentally, this allows us to directly compare in a single experiment both the q-BIC and the corresponding bright mode, hence their frequency separation (approximately 10 cm$^{-1}$, which matches Table 1).

SNOM images of polaritonic are not in general easy to interpret due to the complexity of the coupling phenomena that occur in the tip-sample system. In particular, several scattering processes can occur and they interfere to form the final image. In this case, the opposition in phase in the dipolar modes causes the bright mode to be visible in one of the half of the antenna, and the BIC mode to be visible on the other. In addition, modes on the major and minor axes have the expected frequencies and are consistent with the selected light polarization.

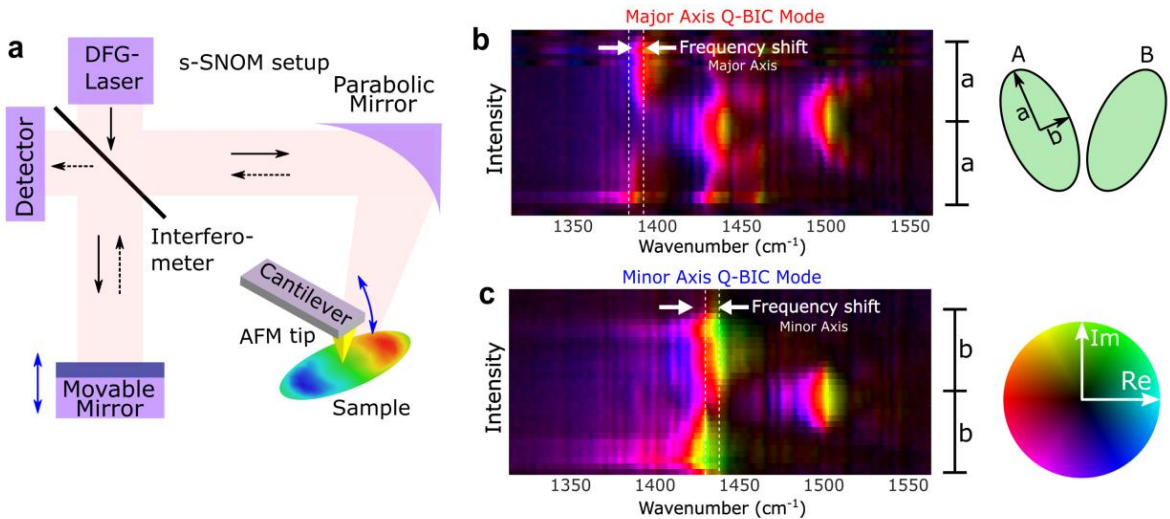

**Figure 5. NSOM characterization. (a)** Schematic of the used nano-FTIR setup. A Michelson interferometer is used to retrieve amplitude and phase of the scattered near-field signal. The system uses a DFG source which guarantees at the same time large spatial coherence and small temporal coherence (resulting in a wideband spectrum). **(b)** nano-FTIR spectra collected as a function of the position of the tip along the major axis of one of the resonators (A), with the polarization of light selected to excite the related q-BIC mode (here, the Da mode of Figure 1d). The bright mode is also visible because it has much stronger coupling with light than the q-BIC one. This allows to measure the shift between the modes []. Other higher order modes are also visible at larger wavenumbers. Color bar: the color represents the phase (from 0 to 360°), brightness the amplitude (normalized to 1). **(c)** Same as **(b)** but for the q-BIC of the minor axis (Db in Figure 1d). To the right of the spectra in **(b-d)** we reported the sizes of the ellipse's axes for comparison.

# *Conclusions*

Our demonstration of BIC and q-BIC states in a polaritonic system paves the way to several new research directions. Since polaritonic resonators can be miniaturized indefinitely, it is possible to create very complex unit cells (even with more than two nanostructures) to create arrays with a very rich spectrum of BIC and dark modes. There are many ways to generalize these results, such as creating more complex patterns where one or more perturbations are introduced along all the dimensions of the array. In all cases, the perturbations can be tuned to control the amount of coupling of these modes with the environment (through the radiative channels), creating interesting trade-offs also for light manipulation, optoelectronics and metasurfaces, besides the obvious applications for sensing and detection. Exploring different materials, such as excitonic and plasmonic monolayers, will make this platform even more flexible thanks to the possibility of tuning these materials electrically.

## *Methods*

**Theoretical Simulations**

Simulated transmittance spectra of the hBN resonators array were obtained using Ansys HFSS finite element method (FEM) electromagnetic solver. The software was used to calculate modes and scattering parameters (S-parameters) of the structures in the frequency domain. The resonator was excited using a port, and a frequency sweep was performed in the range of 1300 cm$^{-1}$ to 1500 cm$^{-1}$ in the mid-IR region.

**Metasurface Nanofabrication:**

Hexagonal boron nitride crystals were grown via precipitation from an Fe solvent in a $N_2/H_2$ atmosphere using isotopically enriched (> 99%) elemental $^{11}$B powder as the boron source material. As the source of nitrogen, we used naturally abundant gas, which is 99.6% $^{14}$N and 0.4% $^{15}$N.

These h$^{11}$BN crystals were mechanically exfoliated via the scotch tape method on a 5 mm thick $CaF_2$ substrate. Suitable flakes were identified using an optical microscope, and their thicknesses were measured with an atomic force microscopy (AFM-Park XE-100). Afterward, a layer of the positive-tone electron beam resist ZEP-520A (2:1 of Anisole: ZEP-520A) was spun onto the $CaF_2$ substrate, baked at 90º C for 180 s and then at 180º C for 180 s. An electrically conductive layer of 10 nm of gold was sputtered with a Kenosistec KS500 confocal sputter coater system on the top of the resist. The patterns were defined using an electron beam lithography (EBL) system (Raith-150 Two) with an acceleration voltage of 30 kV, an aperture size of 20 μm and an area dose of 90 μC/cm$^2$. After the lithography, the conductive gold layer was etched with a commercial gold etchant for 10 seconds, and then the resist was developed in the ZEP Developer (ZED) at 5°C for 60s with a subsequent 30 s rinse in IPA. The patterns were verified with an optical microscope. hBN was then etched with an inductively coupled plasma-reactive ion etching (ICP-RIE) with ZEP as a masking layer. ICP-RIE was performed for 80 s using 10 sccm of $CHF_3$ under the pressure of 5 mTorr with 60 W of RF power and 450 W of ICP power. Finally, the ZEP mask was stripped by placing the sample in remover-PG at 80°C followed by rinsing in the DI water.

**Measurements**

The far-field optical characterization of the hBN resonators was performed using a commercial Fourier transform infrared spectrometer (Thermofisher iS50 FTIR). Transmittance and reflectance were measured in the range 1300 cm$^{-1}$ to 1500 cm$^{-1}$ with an illumination aperture of 20 microns. A holographic polarizer was added in front of the detector for polarization-resolved measurements.

Near-field characterization was performed with a commercial Neaspec system using the related nano-FTIR module with a broadband spatially coherent DFG (difference frequency generation) mid-IR source and a Michelson interferometer using the full travel range of the mirror (1.5 mm, corresponding to a 3 cm$^{-1}$ spectral resolution). Demodulation of the tip-sample scattered signal with a lock-in set at two times the frequency of the AFM tip is performed to select the near-field signal coming from the pure tip-sample interaction (hence, reaching a spatial resolution compared to the tip radius, about 50 nm). Due to the geometry of the interferometer, it is possible to retrieve both the amplitude and the phase of the scattered near-field. A systematic offset of 14 cm$^{-1}$ was observed with respect to the expected response of the materials and the FTIR measurement. We also experienced this offset while we were measuring known reference spectra. In particular, we selectively operated with single wavelengths of a CW Ti:Sa laser, tunable in the 700-1000 nm range, and collected the reflected light when the tip (silicon with a Pt-Ir coating) was on top of a silicon substrate. The shift at all wavelengths oscillates between 0.5 to 1 % of the measurement wavelength. All measurments presented here have been corrected to take into account this offset. Hyperspectral maps with phase and amplitude were obtained by scanning along the axes of the ellipses and in a raster grid.

# *Data availability*

All data supporting the findings of this study are available within the article and its supplementary information.

## *Code availability*

The code that supports the findings of this study is available from the corresponding authors upon reasonable request.

## *Acknowledgements*

We acknowledge the financial support of the European Research Council (ERC) under Grant Agreement No. ERC-2020-STG 948250 (SubNanoOptoDevices). We gratefully acknowledge the IIT cleanroom facility and staff. Support for $h^{11}BN$ crystal growth was provided by the Office of Naval Research, award numberN00014-22-1-2581.

## *Author contributions*

H.G., M.T. developed the theory and performed numerical simulations of the structure. E.J. and J.H.E. synthesized and grew the isotopically pure $h^{11}BN$ crystals. H.G., T.C. fabricated the devices with inputs from A.T. and M.T. H.G., A.T. F.dA. and M.T. performed FTIR measurements. G.V., A.A. performed the near field scans. H.G. and M.T. analysed the data and wrote the paper with inputs and feedback from all the authors. M.T. led the project.

## *Competing interests*

**The authors declare no competing interests.**

## *Additional information*

**Correspondence and requests for materials** should be addressed to Harsh Gupta and Michele Tamagnone